\documentclass[%
 aip,
 amsmath,amssymb,
 reprint,%
]{revtex4-1}

\usepackage{graphicx}
\usepackage{subcaption}
\usepackage{dcolumn}
\usepackage{bm}

\usepackage[utf8]{inputenc}
\usepackage[T1]{fontenc}
\usepackage{mathptmx}
\usepackage{etoolbox}

\makeatletter
\def\@email#1#2{%
 \endgroup
 \patchcmd{\titleblock@produce}
  {\frontmatter@RRAPformat}
  {\frontmatter@RRAPformat{\produce@RRAP{*#1\href{mailto:#2}{#2}}}\frontmatter@RRAPformat}
  {}{}
}%
\makeatother
\begin{document}

\preprint{AIP/123-QED}

\title[]{Effective Phonon Dispersion of $\beta$- (Al$_x$Ga$_{1-x}$)$_2$O$_3$ alloy semiconductor}
\author{Ankit Sharma}
 \author{Animesh Datta}
\author{Uttam Singisetti}%
 \email{uttamsin@buffalo.edu}
\affiliation{ 
Department of Electrical Engineering, University at Buffalo, Buffalo, NY, 14260, USA
}%

\date{\today}

\begin{abstract}
In this work, the Effective Phonon Dispersion of $\beta$- (Al$_x$Ga$_{1-x}$)$_2$O$_3$ alloy is investigated using the Phonon unfolding formalism. To capture the true randomness of the system, supercells are designed using the technique of special quasirandom structues. The phonon unfolding technique is then used to obtain the effective phonon dispersion (EPD) for aluminium fractions of 18.75 \% and 37.5 \% with respect to gallium atoms. The impact of disorder on the high frequency and DC dielectric constant is also investigated. The unfolding procedure was also applied to the dipole
moment where the projection probability was used to obtain the IR spectra of the disordered system. The calculated properties gives us further insights to calculate the transport properties in these alloy semiconductors while capturing the true randomness of the system.
\end{abstract}

\maketitle
\section{\label{sec:level1}Introduction and Overview}
$\beta$-Ga$_2$O$_3$ is a promising Ultra Wide Bandgap (UWBG) semiconductor which has attracted the worldwide attention of researchers due to its highly promising properties. A large band gap of 4.8-4.9 eV \cite{tippins1965optical,orita2000deep,he2006first}, high critical electrical field strength of 8 MV/cm \cite{higashiwaki2013depletion} in addition to high electron saturation velocity \cite{ghosh2017ab} results in  high Baliga's Figure of Merit (BFoM)\cite{higashiwaki2014development,higashiwaki2016recent} making it an ideal candidate for RF, power and space electronics\cite{10318032} applications. Mature bulk crystal growth techniques like Czochralski(CZ), float zone(FZ) \cite{baldini2016si,stepanov2016gallium,galazka2016scaling}and thin film techniques like metal organic chemical vapor deposition (MOCVD), molecular beam epitaxy (MBE), and halide vapor phase epitaxy (HVPE), with controllable doping(10$^{16}$ cm$^{-3}$- 10$^{20}$ cm$^{-3}$) \cite{tang2022high,feng2019mocvd,sasaki2012device}  has accelerated the rapid development of $\beta$-Ga$_2$O$_3$ devices. Reports have shown devices with  multi-kV breakdown, high cut off frequencies and high figure of merit in $\beta$-Ga$_2$O$_3$ devices \cite{bhattacharyya2021multi,bhattacharyya20224,sharma2020field,saha2023scaled,lv2020lateral,9933753}. 
\\
Recently there has been increasing interest in $\beta$- (Al$_x$Ga$_{1-x}$)$_2$O$_3$ which is an alloy of $\beta$-Ga$_2$O$_3$ to grow and fabricate AlGaO/GaO heterostructure devices to utilize the benefits of 2D electron gas (2DEG) formed at the hetero junction. Theoretical calculations have predicted the low field electron mobility of 500 cm$^2$/V-s in AlGaO/GaO heterostructures \cite{kumar2020low} is  higher than the bulk mobility of 200 cm$^2$/V-s at room temperature\cite{oishi2015high,zhang2019mocvd}. Previous reports have demonstrated the successful growth of high quality AlGaO alloys\cite{zhang2019mocvd,anhar2019mocvd,bhuiyan2021mocvd}. HEMT's using AlGaO/GaO heterostructures with high transconductance and cutoff frequencies have already been reported  \cite{vaidya2021enhancement}. There has also been theoretical reports of optical properties, band offset and orientation dependent band offset in AlGaO alloys\cite{mu2020orientation,wang2018band}.
\newline
However to fully utilize the advantages of AlGaO alloys and heterostructures it critical to understand the band structure and phonon properties of the material. In our previous work, the effective band structure of AlGaO alloys was reported capturing the full disorder in the alloy system using the supercell approach\cite{sharma2023effective}. Along with the bandstructure properties, modeling phonon is a very important and significant problem both in terms of understanding thermal properties and also charge transport mechanism in semiconductors. Understanding these properties become even more important for materials used in high power electronics since inefficient extraction of heat from the device may cause thermal failure as these devices operate at very high voltages. 
\newline
Motivated with the results of the electronic structure, the BZ unfolding scheme is applied to calculate the phonon dispersion in disordered systems which is the main focus of this work.  The disordered system was modeled using SQS supercells to capture the effect of local disorder which is lost in the effective medium theories such as the virtual crystal approximation (VCA). The theory of the calculation of phonon spectra is well established for materials that can be modeled using periodic unit cells such as Si, GaAs, GaN, $\beta$-Ga$_2$O$_3$ etc. to name a few. The widely used methods to calculate the phonon dispersion are the finite difference and density functional perturbation theory (DFPT). These approaches when applied to disordered or finite size systems such as alloys, nanostructures suffer the same challenges as that of the electronic structure due to the broken translational symmetry. 
\\
Till now, VCA has been the go to method to investigate both the electronic and phonon properties in alloys such as that used for $In_xGa_{1-x}As$ alloy\cite{lee2013unfolding}. But when VCA is applied for phonons in alloys, the results are expected to be even more inaccurate when compared to the VCA electronic structure. The major reason is that when an atom is substituted to form an alloy such as in the current case of AlGaO where Ga is replaced by Al, the difference in the atomic mass between Al (26.981 a.u.) and Ga (69.723 a.u.) is very large which significantly affects the lattice dynamics when compared to $\beta$-Ga$_2$O$_3$. In the case of electronic structure, the valence electronic configuration of the two atoms is the same since they belong to the same group on the periodic table, as such the electronic properties are not expected to change much compared to the phonon dominated properties. Hence methods beyond VCA are required for accurate modeling of phonons in disordered system. 
\newline
As stated earlier, the electronic band unfolding concepts is extended to phonon in disordered systems that are modeled using supercells under the formalism proposed by Boykin\cite{boykin2014brillouin} which is then used to calculate the effective phonon dispersion (EPD). The technique is further used to calculate the dipole moment and the IR spectra and some other auxiliary properties such as the DC dielectric constant for $\beta-(Al_xGa_{1-x})_2O_3$ which gives us further insights into the transport properties of the system.
\section{Phonon Unfolding}
The fundamental principle underlying the phonon unfolding procedure is equivalent to that of the electronic structure unfolding scheme. Firstly the local disorder is captured using SQS (Special Quasirandom Structures)\cite{zunger1990special,wei1990electronic,wang1998majority} supercells that are generated using the Alloy Theoretic Automated Toolkit (ATAT)\cite{van2013efficient,van2009multicomponent}. The SQS for the 40-atom supercell was generated using the 10-atom PC as the fundamental repeating unit in the $2\times2\times1$ configuration. The generated supercell was subjected to volume and atomic relaxation such that the forces on each atom is less than $1\times10^{-4}Ry/au$. Subsequently the self-consistent calculations were performed using Quantum Espresso\cite{giannozzi2009quantum,giannozzi2017advanced,gonze1997dynamical,gonze1995adiabatic} on a $\Gamma$ centered $3\times3\times6$ reciprocal space $\vec{\textbf{k}}$-point grid with a planewave energy cutoff of 100Ry (1360eV). Then DFPT was used to calculate the phonon on a reciprocal space $\Gamma$ centered $3\times3\times6$ $\vec{\textbf{q}}$-point grid. The motivation behind using the DFPT\cite{gonze1995adiabatic,gonze1997dynamical} approach is to obtain the perturbation in the self-consistent potential as a result of the lattice motion which can then be further used to calculate the short range deformation potential scattering rate. The phonon calculation was done for supercell of size 40-atoms with 18.75\% and 37.5\% Al fraction to model the alloy disorder. Beyond this size, the calculation of the full phonon dispersion becomes computationally very challenging, hence for supercells with more than 40-atoms, the phonon calculation is just limited to the $\Gamma$ point. 
\newline
The $\beta-(Al_xGa_{1-x})_2O_3$  (SC) is generated using the 10-atom $\beta-Ga_2O_3$ primitive cell (PC) as stated earlier. As such they are commensurate in both the real and reciprocal spaces. This geometrical relations between the reciprocal lattices can be mathematically represented in terms of the wavevectors of the PC and he SC as given in Eq.\ref{eq: phunfld1}
\begin{equation}
    \vec{q}_n = \vec{Q} + \vec{G}_n
    \label{eq: phunfld1}
\end{equation}
where $\vec{\textbf{q}}$ represents the PC phonon wavevector, $\vec{\textbf{Q}}$ is the SC phonon wavevector and $\vec{\textbf{G}}$ is the SC reciprocal space translation vector. Due to this geometrical relation between the SC and the PC Brillouin zones, the SC phonon eigenstates can be represented as a linear combination of the PC phonon eigenstates at wavevector $\vec{\textbf{Q}}$ and $\vec{\textbf{q}}$ respectively that satisfy Eq.\ref{eq: phunfld1}. An equivalence between the electron wavefunctions and the phonon polarization vectors as their respective representation of eigenstates can be drawn to arrive at the unfolding relation. Adopting the mathematical formalism of Boykin\cite{boykin2014brillouin}, the phonon eigenvectors are represented as Bloch waves in accordance with the periodic boundary condition as given in Eq.\ref{eq: phbloch}
\begin{equation}
    u_{n,\alpha,w} = b_{(\alpha,w)}e^{i\vec{q}.\vec{\rho_n}}
    \label{eq: phbloch}
\end{equation}
where the PC phonon wavevector $\vec{\textbf{q}}$  is determined from the Born-von Karman periodic boundary condition,  $\alpha$ represents the atom inside the PC and \textbf{w} is the cartesian direction. \textbf{b} is the cell periodic part that modulates the planewave in the Bloch expression of the phonon eigenstate and \textbf{u} represents the phonon polarization vector/eigenstate. The cell periodic part is calculated by diagonalizing the dynamical matrix \textbf{D} which is in turn obtained by the application of the DFPT on the unit cell. In the case of phonons which are defined as the collective oscillation of the atoms, each atom has three degrees of freedom corresponding to the three cartesian directions, hence for a given system with N atoms, there are 3N phonon modes or the fundamental vibrational modes. Beyond this, the scheme proposed by Boykin\cite{boykin2014brillouin} is adopted to obtain the projection probability which is equivalent to the electronic eigenstate projection probability. One of the constraints under this formalism is that the phonon eigenstates be orthonormal.
\newline
Using the orthonormality of the phonon eigenstates, the projection probability is calculated as given in Eq.\ref{eq: u7} \cite{boykin2014brillouin}
\begin{equation}
    P(E_p,\vec{q}_m) = \sum_{\alpha,w}|C_p^{(\alpha,w)}|^2
    \label{eq: u7}
\end{equation}
where $E_p$ is the energy of the SC phonon in the mode \textbf{p}, $\vec{\textbf{q}}$ represents the PC phonon eigenvector, \textbf{C} is the projection magnitude of each PC mode onto a given SC phonon mode. \textbf{P} is the projection probability that gives the probability of the existence of a phonon eigenstate at energy $E_p$ at the wavevector $\vec{q}$. The cumulative probability can then be calculated by summing over all the projection probability upto a given supercell phonon eigenenergy. This cumulative probability is represented as jumps with each jump representing a band crossing as discussed in the results section. The cumulative probability is given in Eq.\ref{eq: u8}. The detailed derivation of the above set of equations is presented in \cite{boykin2014brillouin}.
\begin{equation}
    P_c(E_\lambda,\vec{q}_m) = \sum_{p=1}^{\lambda}P(E_p,\vec{q}_m)
    \label{eq: u8}
\end{equation}
The effective phonon dispersion is calculated from the projection probability by calculating the phonon spectral function which is equivalent to that calculated in the electronic structure unfolding case. The expression for the spectral function is given in Eq.\ref{eq: u9}
\begin{equation}
    A(\epsilon,\vec{q})_m) = \sum_{p=1}^{3rNc}P(E_p,\vec{q}_m)\delta(\epsilon-E_p)
    \label{eq: u9}
\end{equation}
where \textbf{A} is the spectral function, \textbf{p} runs over all the SC modes, $E_p$ is the supercell phonon eigenenergy in mode \textbf{p} and wavevector $\vec{Q}$ and $\epsilon$ is the energy grid with an increment of 0.01eV. The $\delta$ function is approximated as a Gaussian with a standard deviation of 0.0025eV. In the next section, the EPD for $\beta-(Al_{0.18}Ga_{0.82})_2O_3$ \& $\beta-(Al_{0.37}Ga_{0.63})_2O_3$ is discussed in detail and using the projection probability the dipole moment and IR strength is also calculated.
\section{Results and Discussion}
\subsection{EPD of $\beta-(Al_{0.18}Ga_{0.82})_2O_3$ \& $\beta-(Al_{0.37}Ga_{0.63})_2O_3$}
\begin{figure}
    \caption{\small The full phonon dispersion with 120 vibrational modes of the 40-atom $\beta$-AlGaO supercell with 18.75\% Al fraction.}
    \centering
    \includegraphics[width=0.90\linewidth]{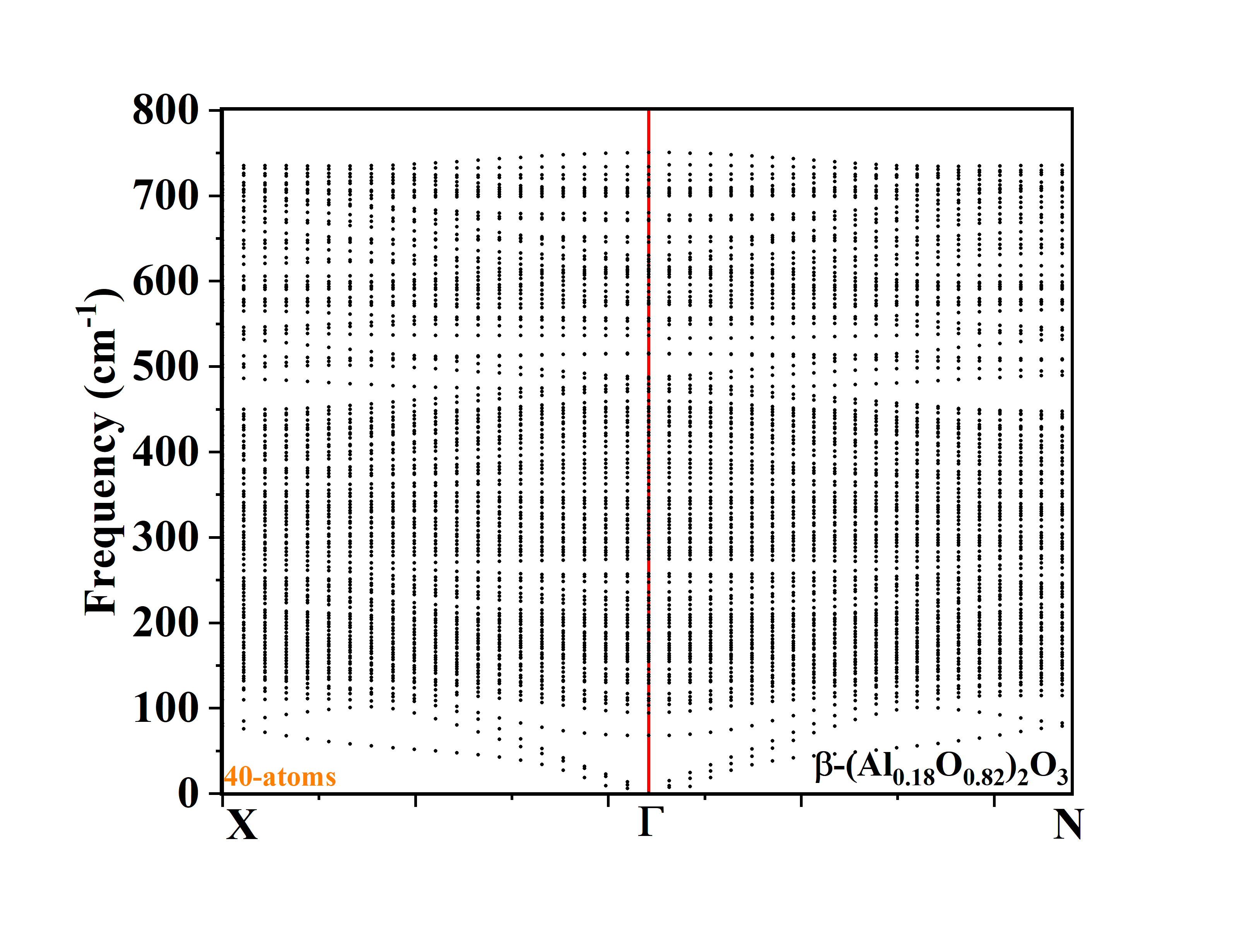}
    \label{fig: 40atmphsc}
\end{figure}

\begin{figure*}[t] 
    \centering
    \subcaptionbox{}{\includegraphics[width=0.45\textwidth]{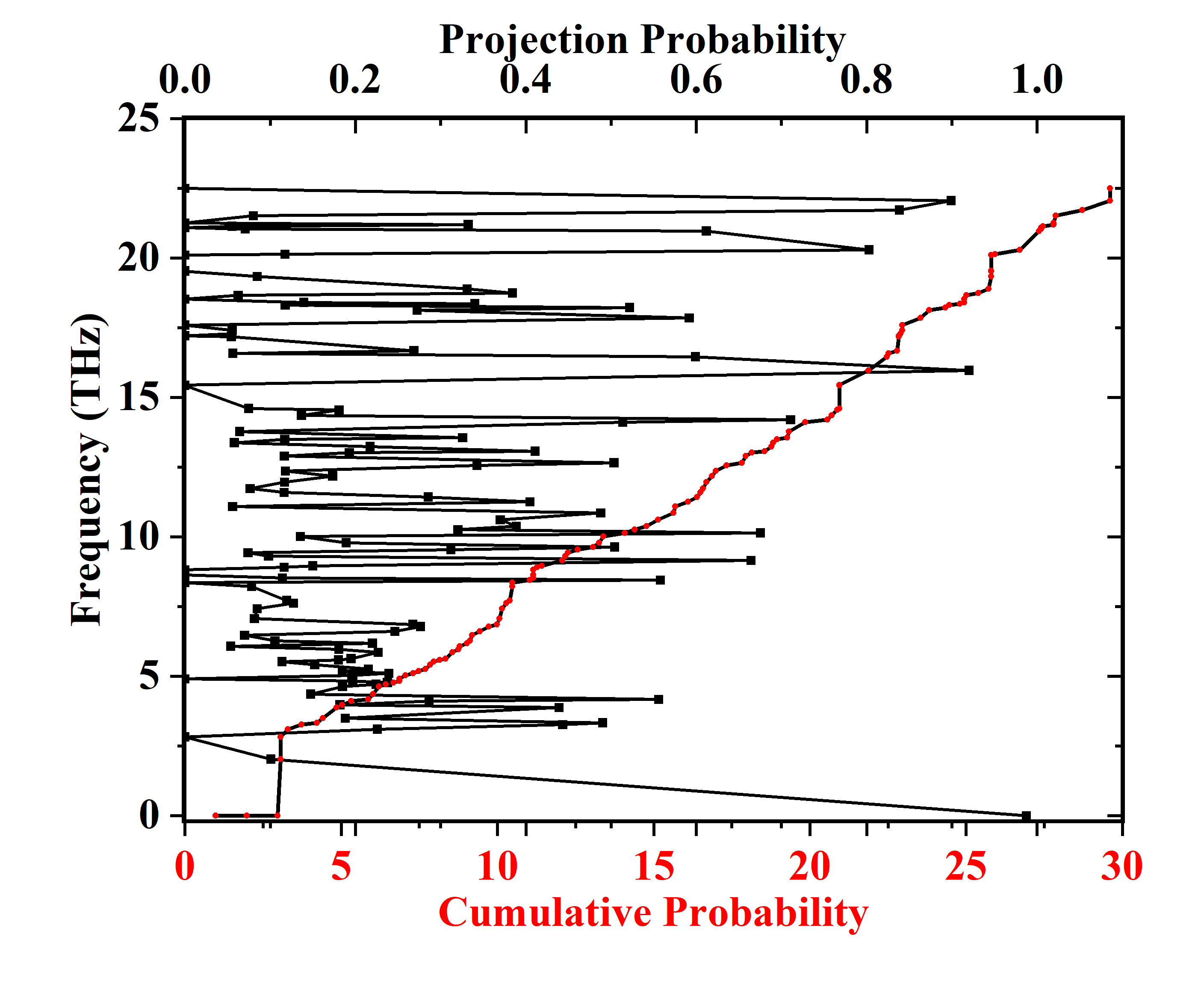}}
    \hfill
    \subcaptionbox{}{\includegraphics[width=0.45\textwidth]{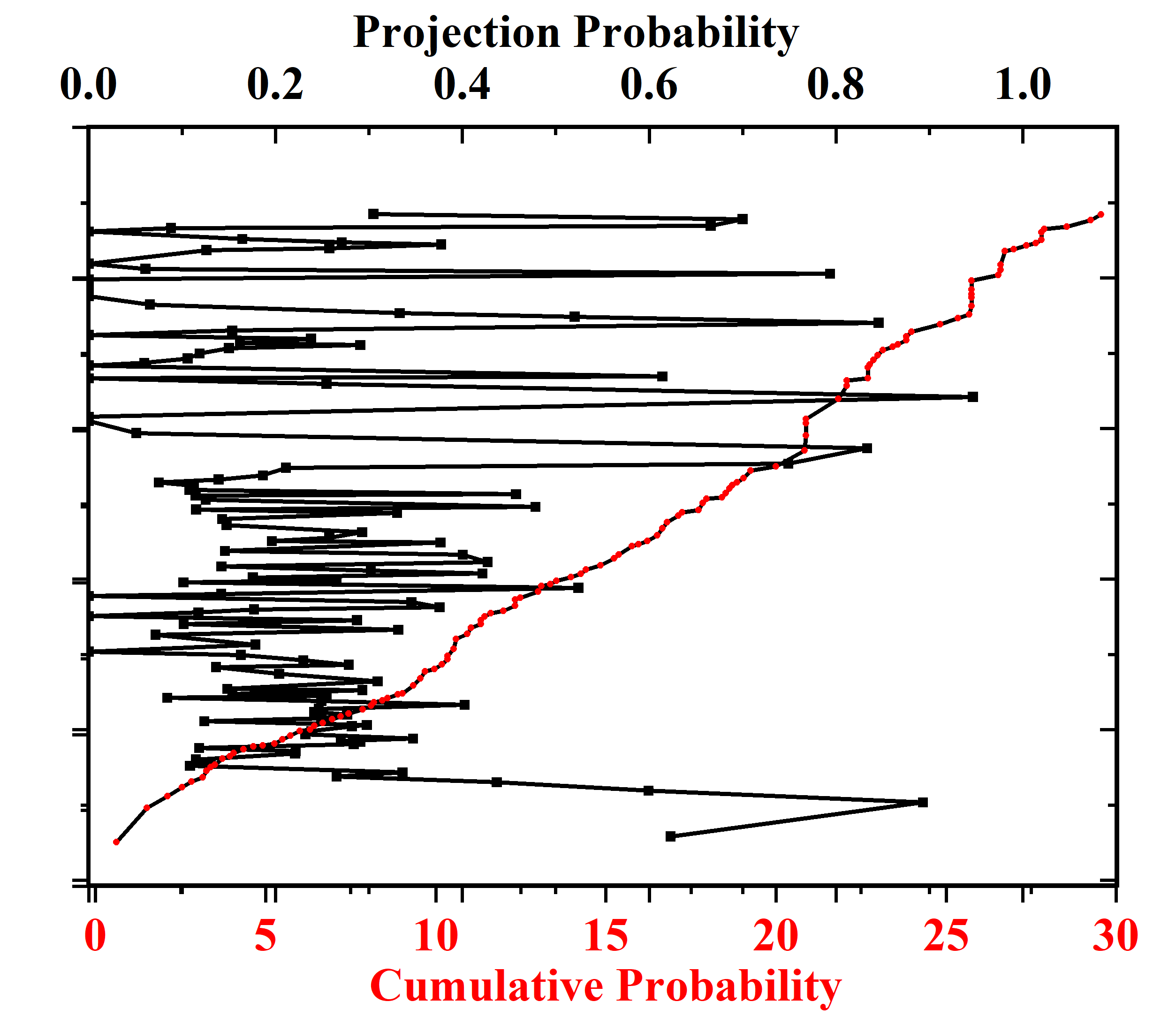}}
    \vspace{0.2em} 
    \subcaptionbox{}{\includegraphics[width=0.45\textwidth]{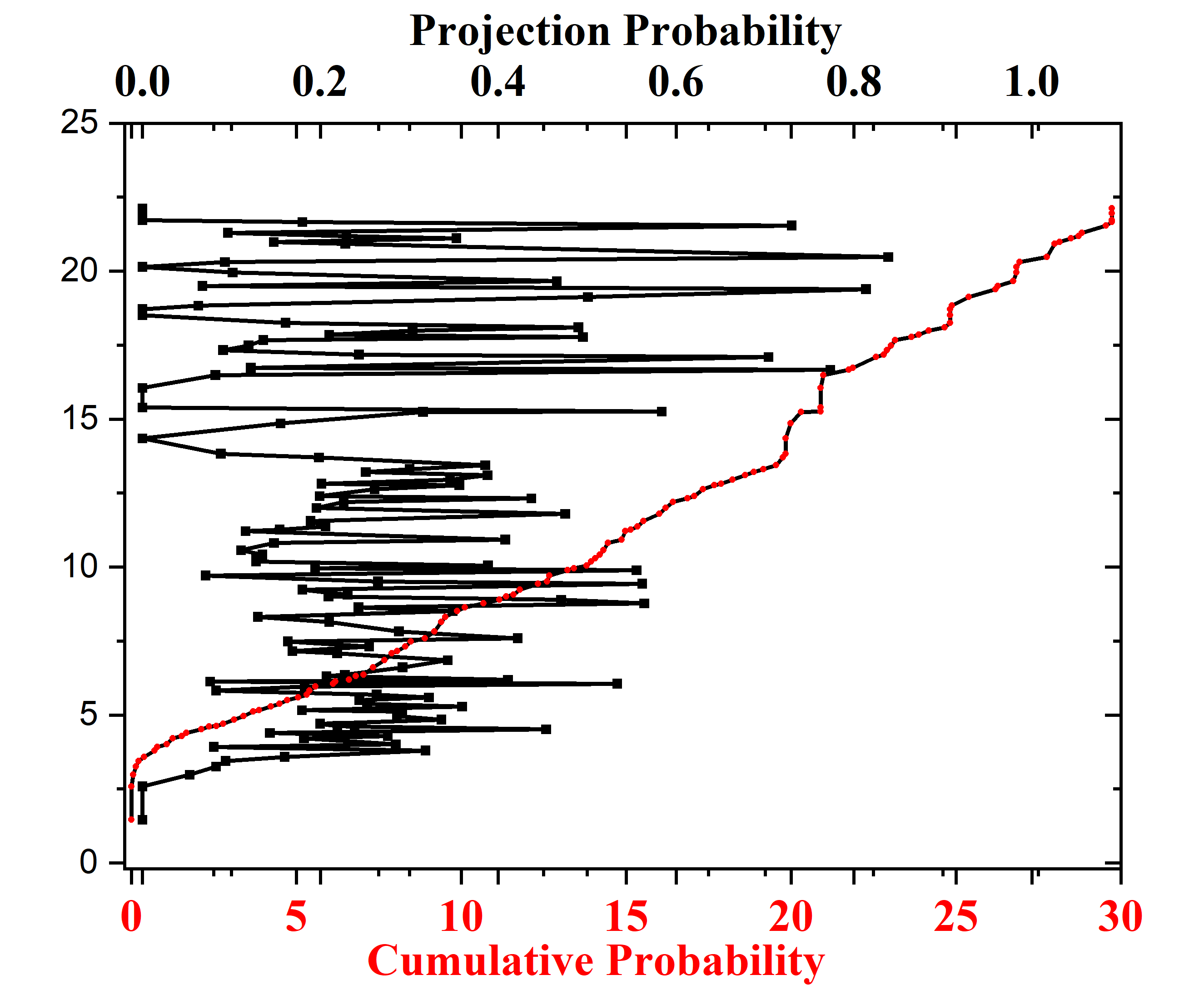}}
    \hfill
    \subcaptionbox{}{\includegraphics[width=0.45\textwidth]{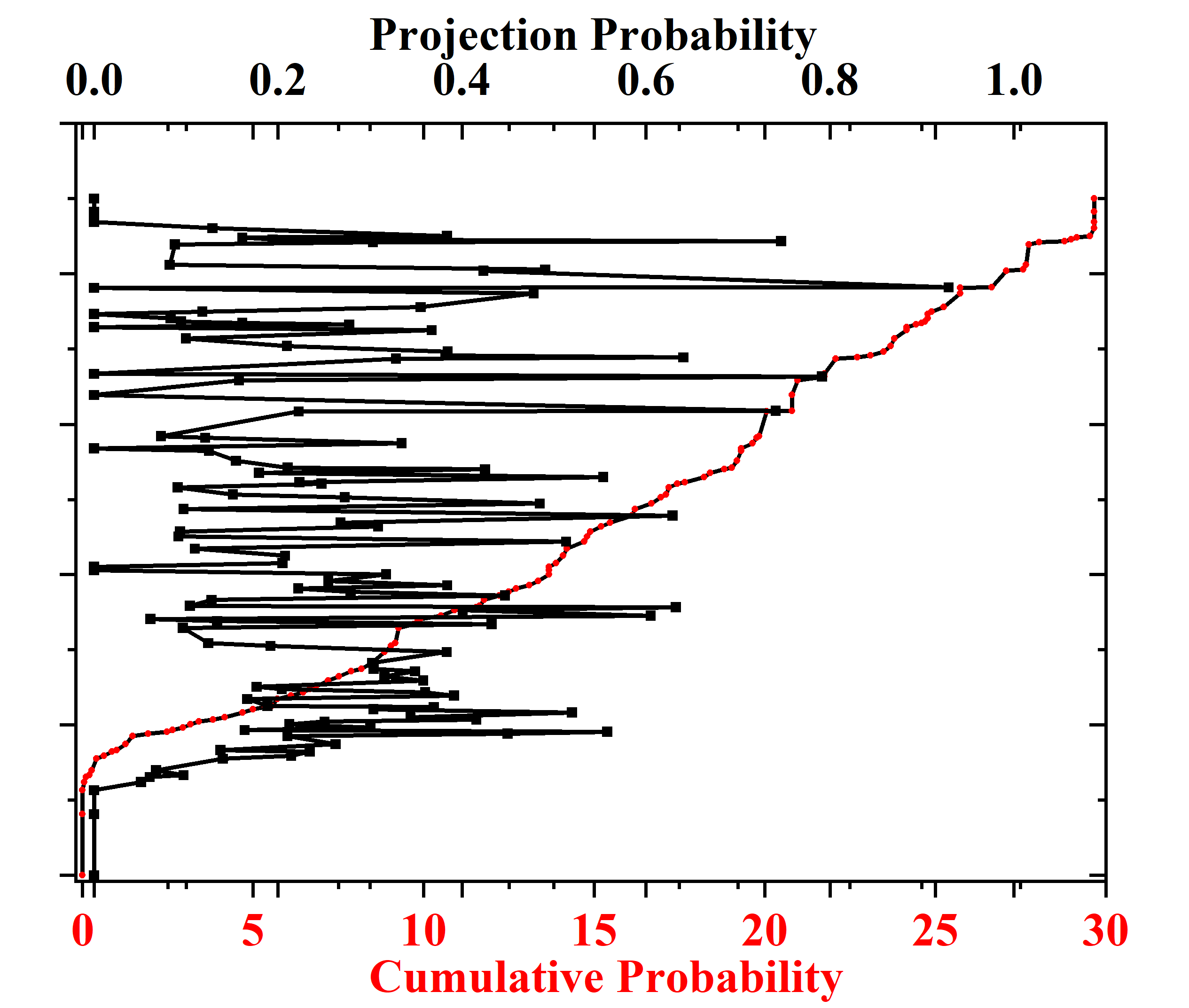}}
    \caption{\small The cumulative probability in red and projection probability for $\Gamma$ point in (a), [0 0.115 0] in (b), [0 0.36 0] in (c) and [0 0.5 0] in (d) in the $\Gamma-N$ direction}
    \label{fig: cprob1}
\end{figure*}

Conventional DFPT calculation of the 40-atom unit cell results in 120 fundamental vibrational modes. Fig.\ref{fig: 40atmphsc} shows the full phonon dispersion as generated by the Fourier interpolation of the dynamical matrix in the irreducible directions of the BZ.
\newline
\begin{figure}[h!]
    \begin{minipage}[c]{0.5\linewidth}
        \centering
        a).
        \includegraphics[width=1.0\linewidth]{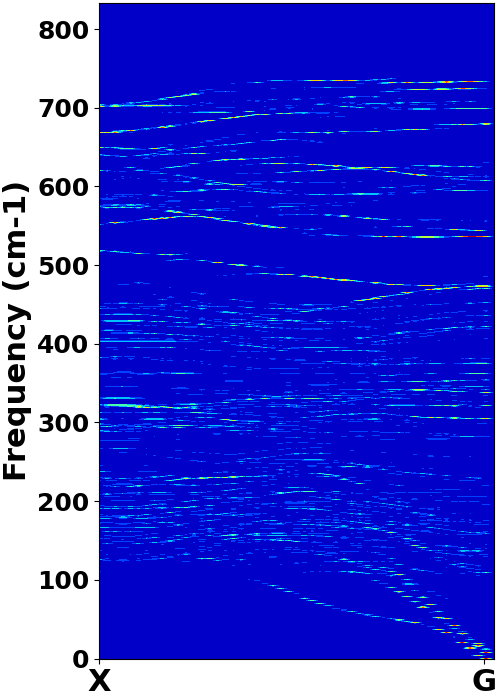}
    \end{minipage}\hfill
    \begin{minipage}[c]{0.5\linewidth}
        \centering
        b).
        \includegraphics[width=1.0\linewidth]{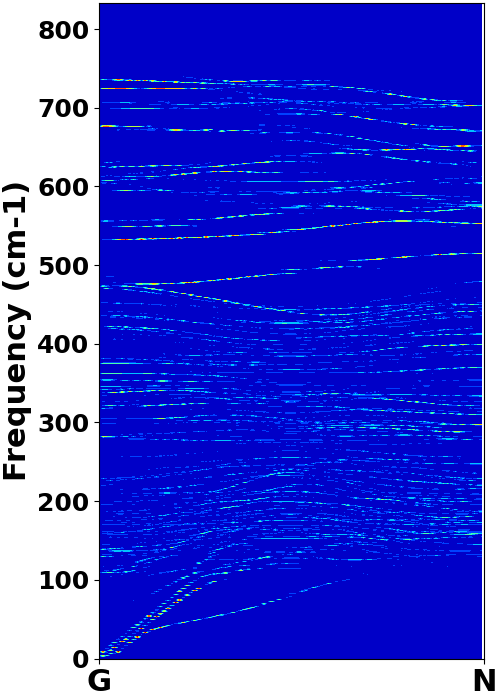}
    \end{minipage}
    \caption{\small The effective phonon dispersion (EPD) for the 40-atom 18.75\% Al fraction (a). in the $X-\Gamma$ direction and (b). in the $\Gamma-N$ direction}
    \label{fig: epd40atm1}
\end{figure}
As can be seen from Fig.\ref{fig: 40atmphsc}, it is very difficult to individually separate each of the phonon modes due to large number of phonon modes clustered within a given energy window. The band unfolding procedure was then applied to obtain the effective phonon dispersion as shown in Fig.\ref{fig: epd40atm1}(a) in the $X-\Gamma$ (X:[0.266 0.266 0]), direction and in Fig.\ref{fig: epd40atm1}(b) in the $\Gamma-N$ (N:[0.0 0.5 0.0]), direction of the PC Brillouin zone. The disorder induces broadening in the effective dispersion as seen in the figure above. For phonons, it is slightly difficult to estimate the band positions since the optical phonons are highly degenerate as can be seen in Fig.\ref{fig: 40atmphsc}, hence the band resolutions is not great. The plot of the cumulative and projection probability for a bunch of $\vec{\textbf{q}}$-point is given in Fig.\ref{fig: cprob1}

  %
       
        
        
\begin{figure}[h!]
    \caption{\small The full phonon dispersion with 120 vibrational modes in the irreducible wedge of the BZ for $\beta-(Al_{0.37}Ga_{0.63})_2O_3$ modeled using 40-atom SQS supercell}
    \centering
    \includegraphics[width=0.9\linewidth]{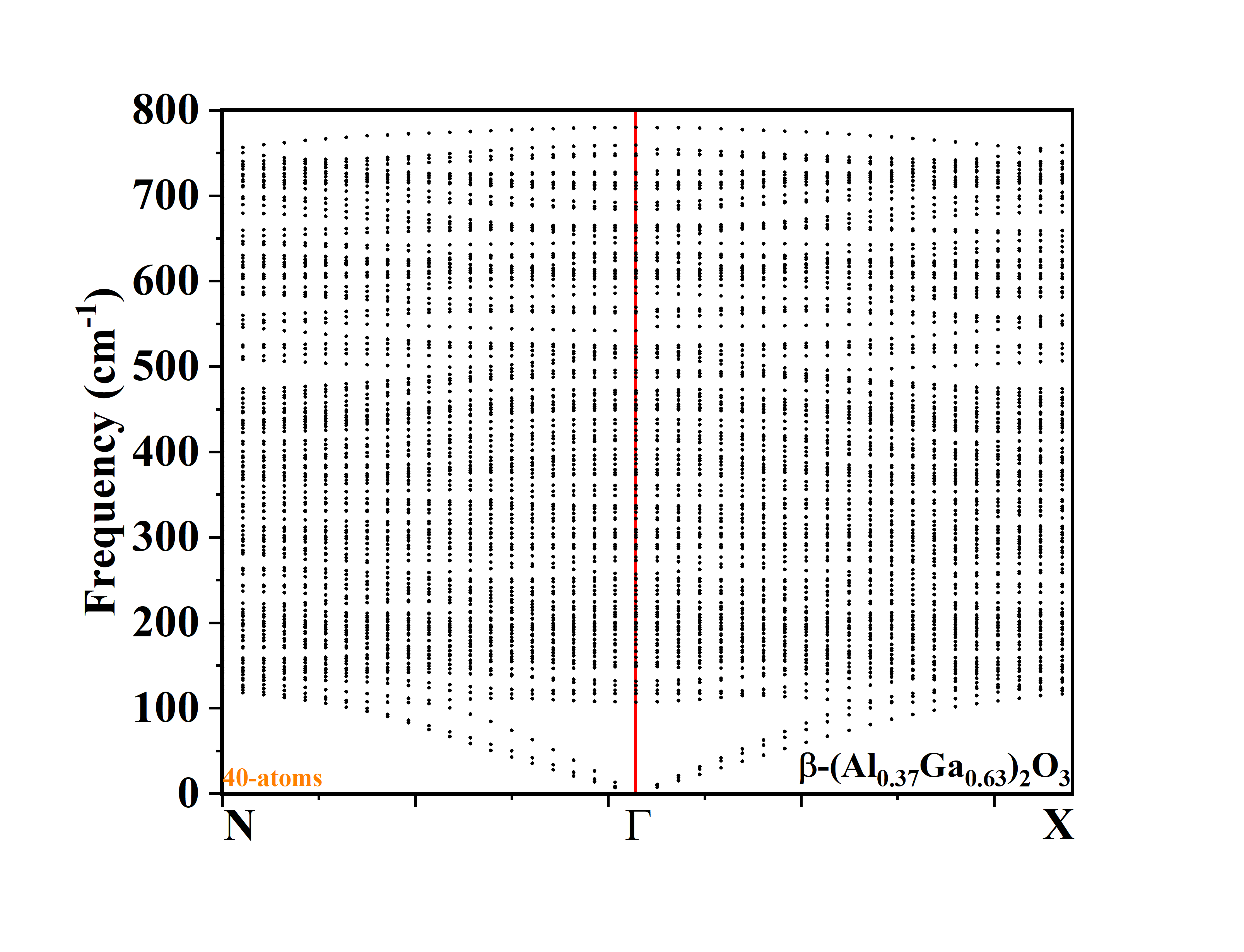}
    \label{fig: 40atomph2}
\end{figure}

As can be seen in Fig.\ref{fig: cprob1}, the cumulative probability in red appears as a series of steps and the projection probability as peaks. Each peak reflects as a step in the cumulative probability. A valid PC band exists whenever the cumulative probability increases by unity. The most challenging part is to identify the valid steps and their resolution because it will determine whether the bands are to be considered isolated or degenerate. This explanation is revisited later while evaluating the IR strength from the unfolded phonon dispersion. Next the full phonon dispersion for $\beta$-AlGaO with 37.5\% Al fraction is presented in Fig.\ref{fig: 40atomph2}
\begin{figure}[h!]
    \begin{minipage}[c]{0.5\linewidth}
        \centering
        a).
        \includegraphics[width=1.0\linewidth]{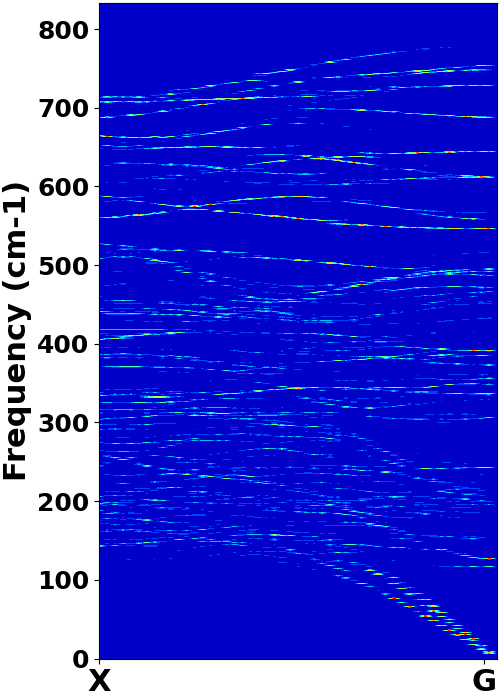}
    \end{minipage}\hfill
    \begin{minipage}[c]{0.5\linewidth}
        \centering
        b).
        \includegraphics[width=1.0\linewidth]{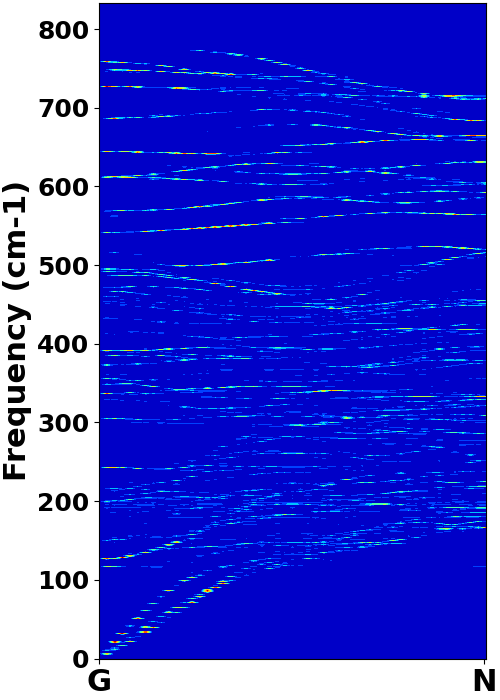}
    \end{minipage}
    \caption{\small The effective phonon dispersion (EPD) for the 40-atom 37.5\% Al fraction (a). in the $X-\Gamma$ direction and (b). in the $\Gamma-N$ direction}
    \label{fig: epd40atm2}
\end{figure}
The band unfolding procedure is applied again to obtain the effective phonon dispersion (EPD) using the phonon spectral function as shown in Fig.\ref{fig: epd40atm2}. As can be seen from Fig.\ref{fig: epd40atm2} that the acoustic modes are stronger compared to optical modes which appear more diffused due to the alloy induced disorder.
From the plot of the cumulative probabilities in Fig.\ref{fig: cprob1}, the maximum value of the cumulative probability is \textbf{30} which is the exact number of phonon modes that is expected from the 10-atom unit cell of $\beta-Ga_2O_3$ which is used to generate the supercell. This is called the sum rule (SR) and acts a check for the unfolding procedure. Hence from the plots it is clear that the 120 phonon modes corresponding to the 40-atom supercell is reduced to 30 phonon modes in the 10-atom PC. 
\section{Dipole Moment \& IR Strength}
\begin{figure}[h!]
    \centering
    \includegraphics[width=1.0\linewidth]{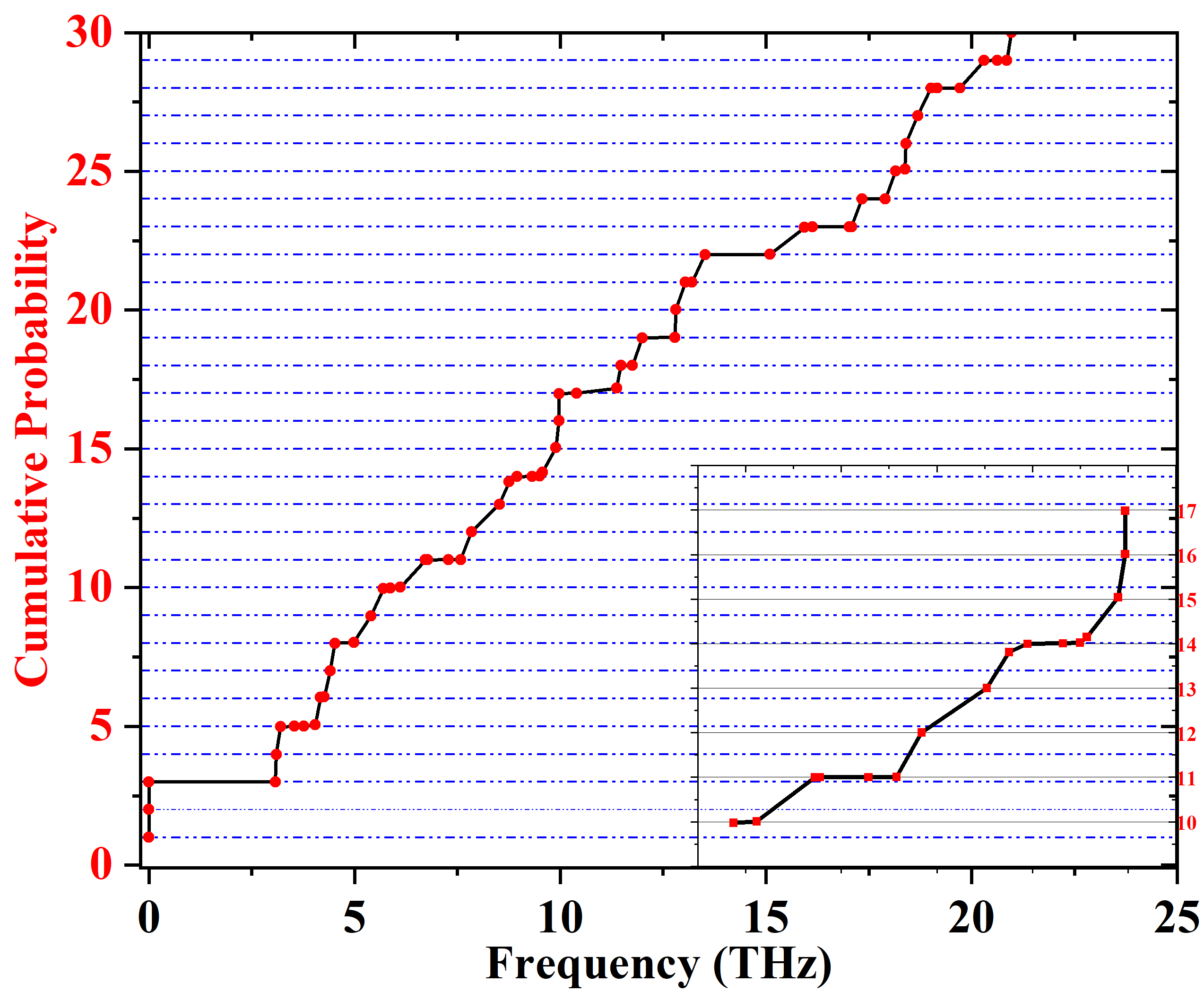}
    \caption{\small The cumulative probability for the 20-atom SC of $Ga_2O_3$ divided into bins bounded by integer crossings. The figure in the inset is of the zoomed in region of the plot}
    \label{fig: 20cprob}
\end{figure}

\begin{figure*}[t] 
    \centering
    \subcaptionbox{}{\includegraphics[width=0.45\textwidth]{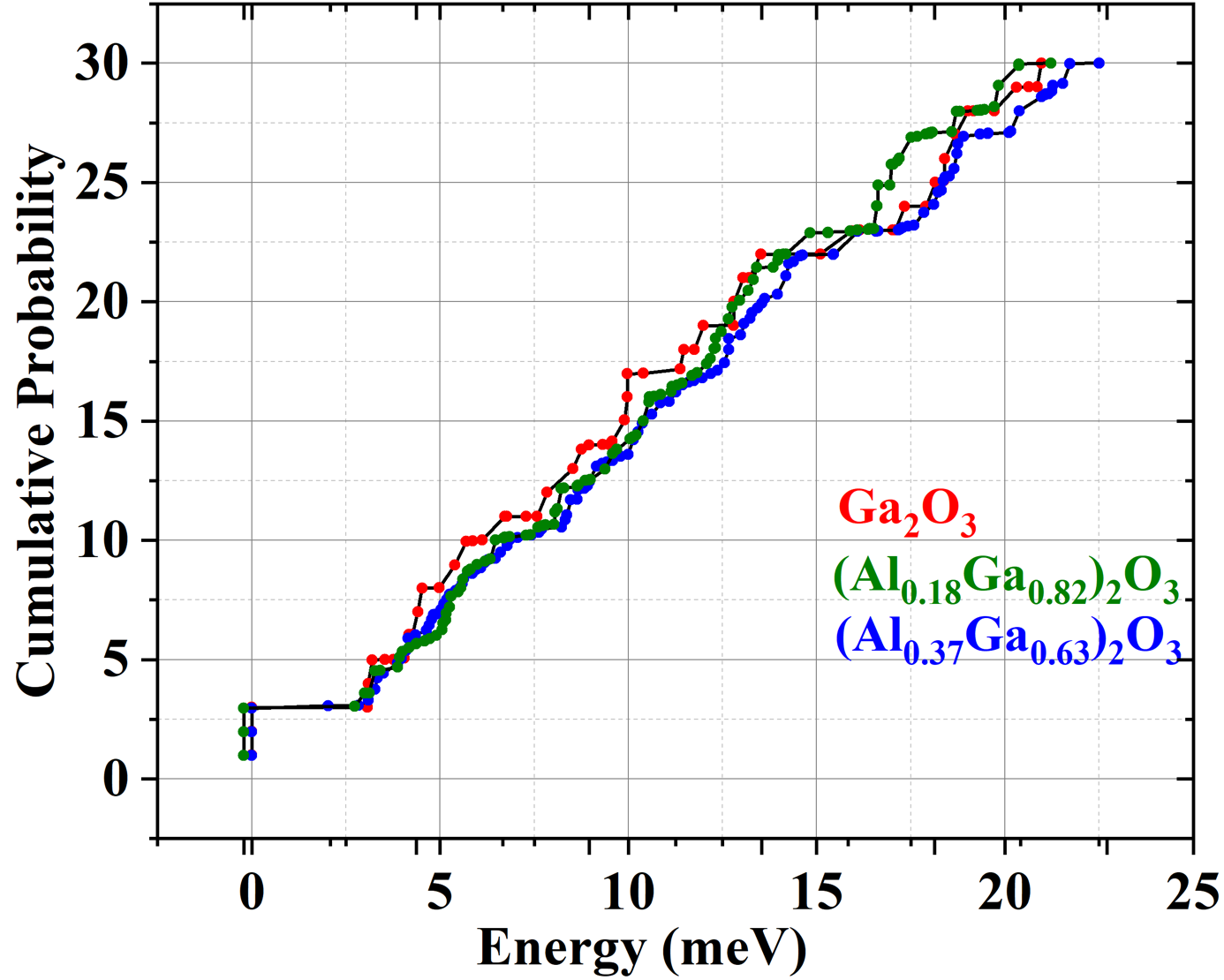}}
    \hfill
    \subcaptionbox{}{\includegraphics[width=0.45\textwidth]{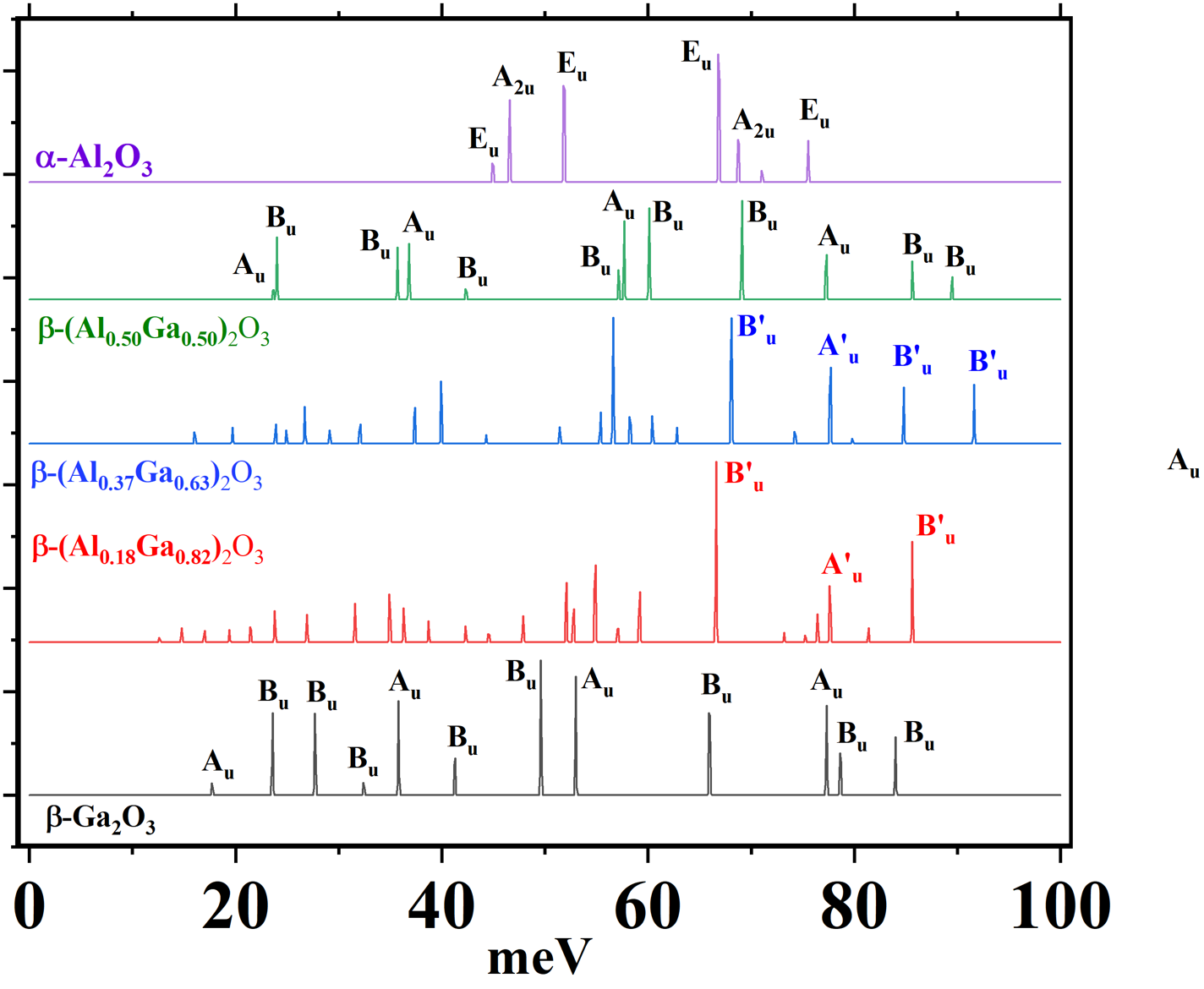}}
    \caption{\small (a) Cumulative Probability used to calculate the energy of the band for $Ga_2O_3$ and $(Al_xGa_{1-x})_2O_3$ (b) Comparison of the the predicted IR spectra from Phonon Unfolding Procedure }
    \label{fig:cpir}
\end{figure*}

For this calculation, the 20-atom SC of $Ga_2O_3$ without any disorder was considered to verify the process of evaluating the dipole moment. The phonon calculation for this 20-atom SC was performed at $\Gamma$ point using DFPT. Also, the Born effective charge and the phonon polarization vectors were calculated under the perturbation theory formalism as implemented in Quantum Espresso\cite{giannozzi2009quantum,giannozzi2017advanced,gonze1997dynamical,gonze1995adiabatic}. This resulted in 60 phonon modes of which 27 modes had non-zero dipole moments i.e. they are IR active modes. The 27 IR modes obtained using the conventional cell is much larger than the experimentally observed 12 IR modes for the pristine 10-atom $\beta$-$Ga_2O_3$ primitive cell. Ideally the number of active modes should be independent of the size and shape of the unit cell used for the calculation. Another factor to account for is the impact of the increased number of IR modes on the polar optical phonon scattering rates which should also be independent of the shape and size. Again motivated by the sum rule which ensures that the reduced number of phonon modes is 30 for $\beta-Ga_2O_3$, the cumulative probability was divided as per integer crossings since each crossing indicates to a presence of a phonon band. Fig.\ref{fig: 20cprob} shows the plot of the cumulative probability at $\Gamma$-point for the 20-atom conventional cell segregated into bins.
\newline
As stated earlier, each integer crossing corresponds to the presence of a phonon band so the SC modes contributing to the PC phonon is identified by dividing the cumulative probability into bins bounded by integers. As such, the dipole moments as given in Eq.\ref{eq: dipm} of each of the SC modes within the integer boundary is calculated and a weighted sum is obtained with the projection probability of each of the SC mode as the weights as given in Eq.\ref{eq: pcdip}.
\begin{equation}
    D_{\nu} = \sum_{\alpha}Z^{*}_{\alpha}u_{\alpha,\nu}
    \label{eq: dipm}
\end{equation}
\begin{equation}
    D^{PC}_{\mu} = \frac{\sum_{\nu=a}^{b}P(E_{\nu})D_{\nu}}{\sum_{\nu=a}^{b}P(E_{\nu})}
    \label{eq: pcdip}
\end{equation}
In Eq.\ref{eq: dipm}, $\nu$ represents the SC phonon mode index, \textbf{Z*} is the Born effective charge for the atom $\alpha$ and \textbf{u} is the associated phonon polarization vector and \textbf{D} is the total dipole moment. In Eq.\ref{eq: pcdip}, \textbf{P} is the projection probability of the supercell phonon onto the primitive cell Bloch space, \textbf{D} is the SC dipole moment and $\mu$ represent the PC phonon mode index. The sum runs between the number of phonon modes that lie within the integer boundaries [\textit{a},\textit{b}]. Using this approach, the 12 IR active modes present in the PC is obtained as given in Table \ref{tab: unfldIR}. The unfolded eigen energies of the IR modes show a close match with the IR modes obtained using the 10 atom PC cell. This provides a validity to this approach of evaluating the IR spectra from the effective phonon dispersion.
\begin{table}[!ht]
    \centering
    \begin{tabular}{c|c|c}
    \hline
    Supercell (IR) & Unfolded (IR) & $\beta-Ga_2O_3$ (IR) \\
    (\textbf{meV}) & (\textbf{meV}) & (\textbf{meV})\\
    \hline
    \hline
        12.72 & ~ & ~ \\
        15.58 & 17.24 & 17.35 \\ 
        17.27 & ~ & ~ \\ 
        17.64 & 22.95 & 22.33 \\ 
        23.57 & ~ & ~ \\ 
        24.28 & 25 & 28.12 \\ 
        27.80 & ~ & ~ \\ 
        31.34 & 31.34 & 30.88 \\ 
        32.43 & ~ & ~ \\ 
        35.25 & 34.52 & 34.23 \\ 
        38.55 & ~ & ~ \\ 
        39.29 & 41.14 & 40.21 \\ 
        41.24 & ~ & ~ \\ 
        43.01 & 47.46 & 47.61 \\ 
        48.63 & ~ & ~ \\ 
        49.60 & 52.9 & 51.8 \\ 
        52.98 & ~ & ~ \\ 
        54.62 & 65.91 & 63.94 \\ 
        65.93 & ~ & ~ \\ 
        66.74 & 76.11 & 76.16 \\ 
        70.66 & ~ & ~ \\ 
        74.04 & 77.97 & 78.24 \\ 
        77.31 & ~ & ~ \\ 
        78.64 & 83.97 & 84.69 \\ 
        79.25 & ~ & ~ \\ 
        83.98 & ~ & ~ \\ 
        85.31 & ~ & ~ \\ \hline
    \end{tabular}
    \caption{\small The comparison of the IR active modes for the 20-atom supercell, the IR activity obtained from the phonon unfolding procedure and that of $Ga_2O_3$ obtained using DFPT. The unfolded IR modes are very close in energy to the modes predicted by the DFPT calculation}
    \label{tab: unfldIR}
\end{table}
The energy of the IR modes was calculated using Eq.\ref{eq: pcenergy} assuming the modes are not degenerate as given in \cite{boykin2014brillouin}.
\begin{equation}
    E^{PC}_{\mu} = \frac{\sum_{\nu=a}^{b}P(E_{\nu})E_{\nu}^{SC}}{\sum_{\nu=a}^{b}P(E_{\nu})}
    \label{eq: pcenergy}
\end{equation}
where $\mu$ is the PC phonon mode index, $\nu$ is the SC phonon mode index, \textbf{P} is the projection probability calculated previously. The sum is run over all the SC modes that fall within the boundary given by integer crossings [\textit{a},\textit{b}] as seen by horizontal lines in Fig.\ref{fig: 20cprob} depicting the cumulative probability. The above mentioned technique was used for $\beta-(Al_xGa_{1-x})_2O_3$ supercells with Al fraction of 18.75\% and 37.5\%. Fig.\ref{fig:cpir}(a) shows the cumulative probablity used to calculate the energy of the bands and Fig.\ref{fig:cpir}(b) shows the predicted IR spectra calculated using the Phonon Unfolding procedure. 

\newpage
\section{LST Relation: $\epsilon^{DC}$}
In this section, the static dielectric constant for different fractions of Al in the AlGaO alloy is calculated. The LST (Lyddane-Sachs-Teller)\cite{lyddane1941polar} relation is used to obtain the DC dielectric constant as given in Eq.\ref{eq: lst1}
\begin{equation}
    \frac{\epsilon^{DC}}{\epsilon^{\infty}} = \Pi_{i}\frac{\omega_{i}^{LO}}{\omega_{i}^{TO}}
    \label{eq: lst1}
\end{equation}
where the product is run over all the modes that have a non-zero dipole moment at the $\Gamma$ point, $\omega$ is the phonon frequency, \textbf{LO} and \textbf{TO} correspond to the longitudinal optical and transverse optical phonon polarization and $\epsilon^{\infty}$ is the high frequency dielectric constant due to response of the valence electrons. For this calculation, the phonon eigenenergies and polarization vectors are only calculated at the BZ center i.e. $\Gamma$ point. For the completeness of this analysis, the calculation is performed for $\beta-Ga_2O_3$, $\beta-(Al_{0.18}Ga_{0.82})_2O_3$,
$\beta-(Al_{0.37}Ga_{0.63})_2O_3$,$\beta-(Al_{0.5}Ga_{0.5})_2O_3$ and the corundum phase of $\alpha-Al_2O_3$ that represents the other extremity of the alloy fraction when all the Ga atoms are replaced by Al. Here corundum is considered to account for the phase transition the AlGaO alloy undergoes when the Al fraction exceed 70\%.\\
\newline
For polar semiconductors, there exist a dipole field due to the electronegativity difference which under the long wavelength limit couples with the phonon modes and results in the LO-TO split. To account for this coupling of the polarization field, a non-analytic term is added to the dynamical matrix in the 3 cartesian directions before its diagonalization. The non-analytic term in added separately in the 3 cartesian directions to incorporate anisotropy in the phonon polarization. Table.\ref{tab: bgaodc} provides the LO-TO energy for $\beta$-Ga$_2$O$_3$ along with the DC and high frequency dielectric constant. 
\begin{table}[h!]
    \centering
    \begin{tabular}{| c | c | c | c | c | c |}
        \multicolumn{2}{c}{\textbf{X-Direction}}&\multicolumn{2}{c}{\textbf{Y-Direction}} & \multicolumn{2}{c}{\textbf{Z-Direction}}  \\ 
        \hline
        \hline
         \textbf{TO} (meV) & \textbf{LO} (meV) & \textbf{TO} (meV) & \textbf{LO} (meV) & \textbf{TO} (meV) & \textbf{LO} (meV)\\
         \hline
         22.30 & 22.40 & & & 22.34 & 27.98\\
         28.12 & 30.80 & 17.35 & 17.53 & 28.12 & 28.83 \\
         30.88 & 32.50 & & & 30.88 & 31.04\\
         40.21 & 41.50 & 34.23 & 39.22 & 40.21 & 40.31\\
         47.62 & 54.50 & & & 47.62 & 53.06\\
         63.94 & 71.60 & 51.80 & 63.73 & 63.94 & 74.35\\
         78.25 & 83.62 & & & 78.25 & 80.04\\
         84.28 & 86.43 & 76.16 & 87.17 & 84.28 & 89.68\\
         \hline
         \multicolumn{2}{c}{$\epsilon^{\infty}$ = 4.39}  & \multicolumn{2}{c}{$\epsilon^{\infty}$ = 4.54} & \multicolumn{2}{c}{$\epsilon^{\infty}$ = 4.55}\\
         \multicolumn{2}{c}{$\epsilon^{DC}$ = 12.49} & \multicolumn{2}{c}{$\epsilon^{DC}$ = 12.05} & \multicolumn{2}{c}{$\epsilon^{DC}$ = 15.14}\\
         \hline
    \end{tabular}
    \caption{\small The LO-TO split and the anisotropic $\epsilon^{DC}$ dielectric constant for $\beta-Ga_2O_3$ where $\epsilon^{\infty}$ is obtained from the DFPT calculation}
    \label{tab: bgaodc}
\end{table}
\begin{table}[h!]
    \centering
    \begin{tabular}{| c | c | c | c | c | c |}
        \multicolumn{2}{c}{\textbf{X-Direction}}&\multicolumn{2}{c}{\textbf{Y-Direction}} & \multicolumn{2}{c}{\textbf{Z-Direction}}  \\ 
        \hline
        \hline
         \textbf{TO} (meV) & \textbf{LO} (meV) & \textbf{TO} (meV) & \textbf{LO} (meV) & \textbf{TO} (meV) & \textbf{LO} (meV)\\
         \hline
         24.01 & 25.41 & & & 24.01 & 29.58\\
         35.70 & 39.04 & 23.65 & 23.69 & 35.70 & 36.41 \\
         42.34 & 43.02 & & & 42.34 & 42.34\\
         57.14 & 57.18 & 36.82 & 41.60 & 57.14 & 57.77\\
         60.11 & 62.66 & & & 60.11 & 65.66\\
         69.03 & 81.10 & 57.67 & 70.33 & 69.03 & 83.10\\
         85.63 & 89.41 & & & 85.63 & 86.89\\
         89.46 & 98.68 & 77.26 & 94.80 & 89.46 & 97.19\\
         \hline
         \multicolumn{2}{c}{$\epsilon^{\infty}$ = 3.62}  & \multicolumn{2}{c}{$\epsilon^{\infty}$ = 3.71} & \multicolumn{2}{c}{$\epsilon^{\infty}$ = 3.80}\\
         \multicolumn{2}{c}{$\epsilon^{DC}$ = 9.9} & \multicolumn{2}{c}{$\epsilon^{DC}$ = 10.67} & \multicolumn{2}{c}{$\epsilon^{DC}$ = 12.25}\\
         \hline
    \end{tabular}
    \caption{\small The LO-TO split and the anisotropic $\epsilon^{DC}$ dielectric constant for $\beta-(Al_{0.5}Ga_{0.5})_2O_3$ where $\epsilon^{\infty}$ is obtained from the DFPT calculation}
    \label{tab: balgaodc}
\end{table}

\begin{figure*}[t] 
    \centering
    \subcaptionbox{}{\includegraphics[width=0.4\textwidth]{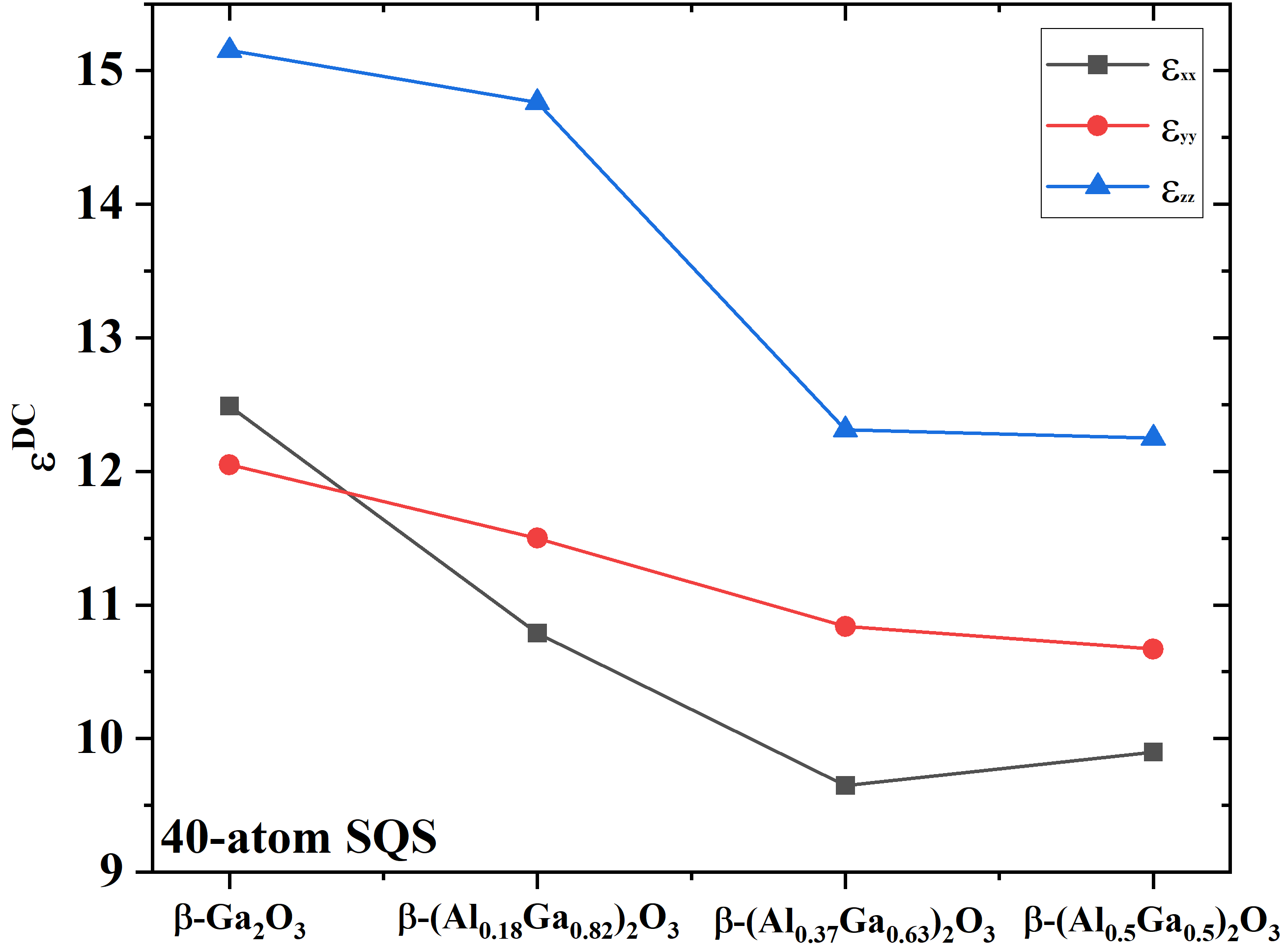}}
    \hfill
    \subcaptionbox{}{\includegraphics[width=0.4\textwidth]{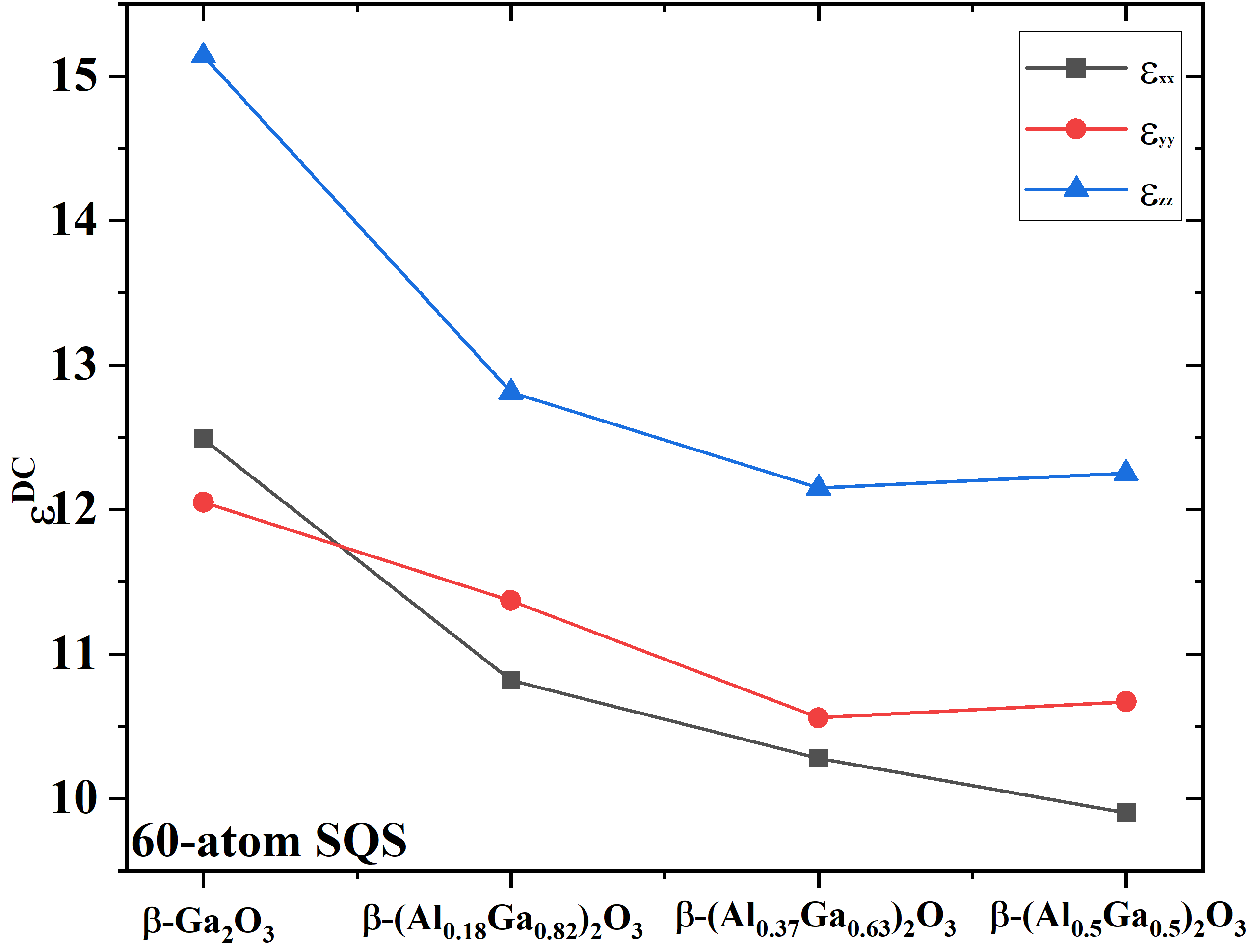}}
    \caption{\small The variation of the $\epsilon^{DC}$ with respect to the Al fraction evaluated using (a). 40-atoms SQS and (b). 60-atoms SQS }
   \label{fig: dcvary}
\end{figure*}
\begin{figure*}[t] 
    \centering
    \subcaptionbox{}{\includegraphics[width=0.45\textwidth]{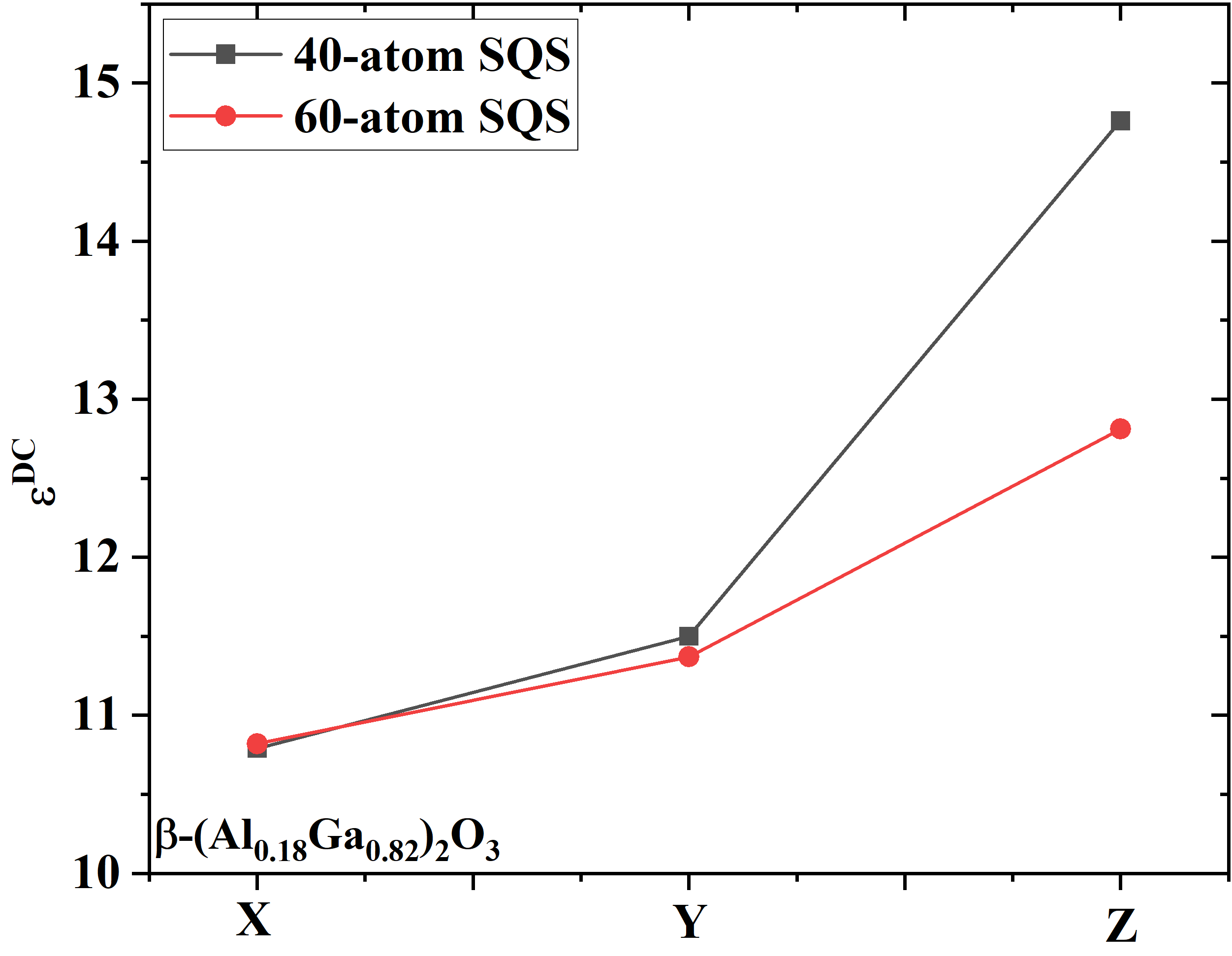}}
    \hfill
    \subcaptionbox{}{\includegraphics[width=0.45\textwidth]{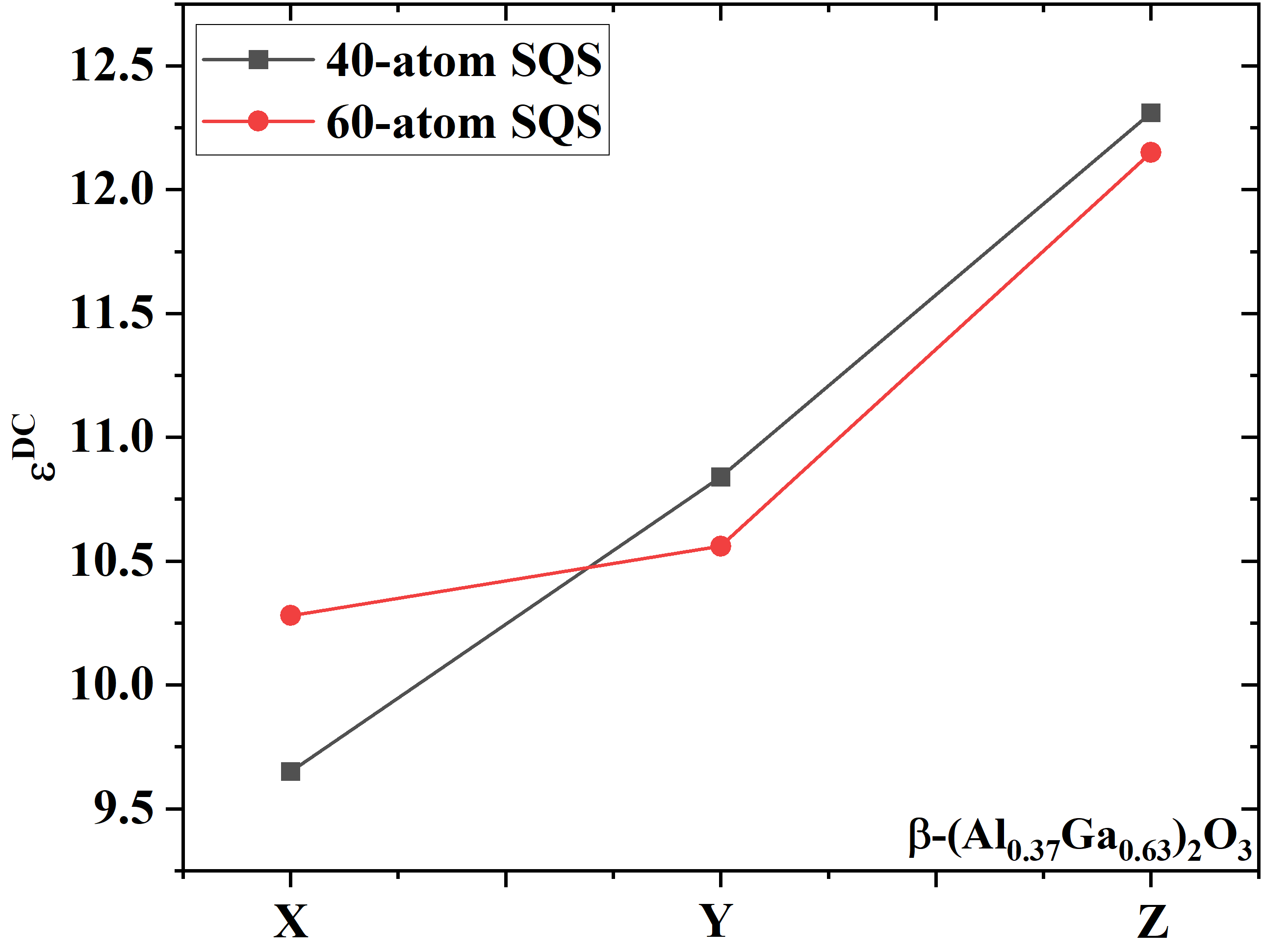}}
    \caption{\small The effect of the size on the DC dielectric constant in the 3 cartesian direction calculated using 40-atoms and 60-atoms supercell for (a). $\beta-(Al_{0.18}Ga_{0.82})_2O_3$ and in (b). $\beta-(Al_{0.37}Ga_{0.63})_2O_3$}
  \label{fig: sizedc}
\end{figure*}

\begin{table}[h!]
    \centering
    \begin{tabular}{| c | c | c | c | c | c |}
         \multicolumn{2}{c}{\textbf{X-Direction}}&\multicolumn{2}{c}{\textbf{Y-Direction}} & \multicolumn{2}{c}{\textbf{Z-Direction}}  \\
         \hline
         \hline
         \textbf{TO} (meV) & \textbf{LO} (meV) & \textbf{TO} (meV) & \textbf{LO} (meV) & \textbf{TO} (meV) & \textbf{LO} (meV)\\
         \hline
         44.94 & 45.09 & 44.94 & 45.09 & 46.56 & 61.24\\
         51.85 & 57.31 & 51.85 & 57.31 & &\\
         66.84 & 75.12 & 66.84 & 75.12 & 68.74 & 103.17\\
         75.52 & 106.77 & 75.52 & 106.77 & & \\
         \hline
         \multicolumn{2}{c}{$\epsilon^{\infty}$ = 3.25}  & \multicolumn{2}{c}{$\epsilon^{\infty}$ = 3.25} & \multicolumn{2}{c}{$\epsilon^{\infty}$ = 3.22}\\
         \multicolumn{2}{c}{$\epsilon^{DC}$ = 10.12} & \multicolumn{2}{c}{$\epsilon^{DC}$ = 10.12} & \multicolumn{2}{c}{$\epsilon^{DC}$ = 12.54}\\
         \hline
    \end{tabular}
    \caption{\small The LO-TO split and the anisotropic $\epsilon^{DC}$ dielectric constant for $\alpha-Al_2O_3$ where $\epsilon^{\infty}$ is obtained from the DFPT calculation}
    \label{tab: alodc}
\end{table}
Table.\ref{tab: balgaodc} and Table.\ref{tab: alodc} provides the LO-TO split and the DC dielectric constant for $\beta-(Al_{0.5}Ga_{0.5})_2O_3$ (periodic structure) and $\alpha-Al_2O_3$ obtained from the LST relation given in Eq.\ref{eq: lst1}. Table.\ref{tab: dcdielectric} provides the $\epsilon^{DC}$ for Al fractions of 18.75\% and 37.5\% modeled using 40-atoms and 60-atoms SQS supercells obtained using the LST relation.
\begin{table}[ht!]
    \centering
    \begin{tabular}{| c | c | c | c | c |}
        \hline
         \textbf{Direction} & 40-atoms & 40-atoms & 60-atoms & 60-atoms  \\
         & 18.75\% & 37.5\% & 18.75\% & 37.5\% \\
         \hline
         \hline
         $\epsilon^{DC}_{xx}$ & 10.79 & 9.65 & 10.82 & 10.28\\
         $\epsilon^{DC}_{yy}$ & 11.50 & 10.84 & 11.37 & 10.56 \\
         $\epsilon^{DC}_{zz}$ & 14.76 & 12.31 & 12.81 & 12.15\\
         \hline
    \end{tabular}
    \caption{\small The DC dielectric constant for 18.75\% and 37.5\% Al fraction obtained using 40-atoms and 60-atoms SQS supercell}
    \label{tab: dcdielectric}
\end{table}
As can be seen from Table.\ref{tab: bgaodc}, \ref{tab: balgaodc}, the phonon modes are polarized in the \textbf{xz} plane and in the \textbf{y} direction for $\beta-Ga_2O_3$ which is well established observation. Similarly polarizations of the phonon eigenvectors are predicted for $\beta-(Al_{0.5}Ga_{0.5})_2O_3$ which is a periodic crystal since all the octahedral gallium sites are preferably occupied by aluminum respectively whereas for $\alpha-Al_2O_3$ the phonon modes polarized in the \textbf{x} and \textbf{y} directions are degenerate and the non-degenerate phonon mode is polarized in the \textbf{z} cartesian direction.

The plot in Fig.\ref{fig: dcvary}(a) \& (b), shows that with increasing Al fraction, the DC dielectric constant decreases indicating that the polar optical scattering will increase due to reduced screening by the valence electrons. Since Al atoms are smaller, the electrons are closer to the nucleus thus the response of the valence electrons to an external perturbation is weaker compared to Ga. As such, with increasing Al fraction in the alloy, the DC dielectric constant decreases. The plot representing the effect of supercell size on the dielectric constant in given in Fig.\ref{fig: sizedc}
 The convergence in the dielectric constant with respect to the supercell size shows that for 18.75\% Al fraction in Fig.\ref{fig: sizedc}(a), when the SC size is increased from 40 to 60 atoms, the \textbf{x} and \textbf{y} components of the dielectric constant does not change whereas their is significant mismatch in the cartesian \textbf{z} component. This can be due to the fact that their is a very strong anisotropy in the dielectric constants for the alloys as can be seen from Table.\ref{tab: dcdielectric}. For 37.5\% Al fraction in Fig.\ref{fig: sizedc}(b), the dielectric constant is not converged in the cartesian \textbf{x} directions whereas, the values show minor variations in the cartesian \textbf{y} and \textbf{z} directions. Supercells with a few more atoms might be required to establish a trend in the convergence of $\epsilon^{DC}$ with the SQS size. Currently due to the computational limitation, the maximum SQS containing 60-atoms is considered for the calculation of the above mentioned quantities. All these results can be further used to calculate the polar optical scattering.
\section{Conclusion}
The effective phonon dispersion of the AlGaO alloy with 18.75\% and 37.5\% Al fraction was obtained using 40 and 60 atoms SQS supercells under the phonon unfolding formalism by Boykin\cite{boykin2014brillouin} which was computationally implemented by an in-house developed code. Our calculations efficiently identify the 30 phonon modes in disordered AlGaO alloys which can be used to study transport properties in these alloy semiconductors. Using the unfolding mechanism, the IR active modes of the AlGaO alloys were also identified which helps in understanding the scattering mechanisms in the material. The LST relation was applied to understand the effect of alloy on the DC dielectric constant which can give further insights on the polar optical scattering mechanism.

\section{Acknowledgments}
We acknowledge the support from the AFOSR (Air Force Office of Scientific Research), under Award No. FA9550-18-1-0479 (Program Manager: Ali Sayir), from the NSF, under Awards ECCS 2019749, 2231026 and the Center for Computational Research (CCR) at the University at Buffalo.

\bibliography{aipsamp}

\end{document}